\documentclass[twoside]{article}

\usepackage{qic,epsfig}
\usepackage[utf8]{inputenc}
\usepackage{graphicx}
\usepackage{subfigure}
\usepackage{epsfig}
\usepackage{amsfonts}
\usepackage{amssymb,amsmath}
\usepackage{verbatim}
\usepackage{bbold}
\usepackage{footnote}
\usepackage{url}
\usepackage{cite}
\usepackage{multirow}
\usepackage{rotating}

\begin{document}
\setlength{\textheight}{8.0truein}    

\runninghead{Fault-tolerant conversion between stabilizer codes by Clifford operations}
            {Yongsoo Hwang et al.}

\normalsize\textlineskip
\thispagestyle{empty}
\setcounter{page}{1}

\copyrightheading{0}{0}{2003}{000--000}

\vspace*{0.88truein}

\alphfootnote

\fpage{1}

\centerline{\bf
FAULT-TOLERANT CONVERSION BETWEEN STABILIZER CODES}
\vspace*{0.035truein}
\centerline{\bf BY CLIFFORD OPERATIONS
}
\vspace*{0.37truein}
\centerline{\footnotesize
Yongsoo Hwang$^{\dag}$,~Byung-Soo Choi$^{\ast}$,~Young-chai Ko$^{\dag}$~and~Jun Heo$^{\dag}$\footnote{junheo@korea.ac.kr}}
\vspace*{0.015truein}
\centerline{\footnotesize\it $^{\dag}$School of Electrical Engineering, Korea University, Seoul, Republic of Korea}
\centerline{\footnotesize\it $^{\ast}$Department of Applied Physics, University of Tokyo, Tokyo, Japan}
\baselineskip=10pt
\vspace*{0.225truein}

\publisher{(received date)}{(revised date)}

\vspace*{0.21truein}

\abstracts{
We propose a scheme that converts a stabilizer code into another stabilizer code in a fault tolerant manner.
The scheme first puts both codes in specific forms, and proceeds the conversion from a source code to a target code by applying Clifford gates.
The Clifford gates are chosen from the comparisons between both codes.
The fault tolerance of the conversion is guaranteed by quantum error correction in every step during the entire conversion process.
As examples, we show three conversions: the $[[5,1,3]]$ stabilizer code and Steane code, Steane code and $[[15,1,3]]$ Reed-Muller code, and Steane code and $(3,4)$-QPC code.
}{}{}

\vspace*{10pt}

\vspace*{3pt}
\communicate{to be filled by the Editorial}

\vspace*{1pt}\textlineskip    
\section{Introduction}\label{sec:introduction}

Thanks to the quantum coding theory and the fault-tolerant quantum computing protocol, it is possible to preserve and manipulate quantum states reliably for a long time~\cite{Gottesman:1997ub,Gottesman:1998gt,Knill:1998bl,Preskill:1998gf,Nielsen:2000vn,Lidar:2013ts}.
Quantum noise is not a fundamental obstacle to build a practical quantum computer any more.
Nevertheless, a quantum computing reliable from quantum noise is very difficult and requires very complicated procedures to initialize quantum states, to run a quantum algorithm on quantum states, to measure quantum states and to protect quantum states from quantum noise during the computing~\cite{Gottesman:1997ub,Gottesman:1998gt,Knill:1998bl,Preskill:1998gf,Lidar:2013ts,Nielsen:2000vn}.
Even preparing ancilla states fault tolerantly is not a simple task.

A fault-tolerant quantum computing begins a computing task by encoding quantum states with a quantum error-correcting code.
The computing then manipulates, transmits and temporarily stores the encoded quantum states, and finally reads out an answer by using various components of a quantum computer.
All the components have to be operated in a fault-tolerant manner that is a quantum computing protocol designed to protect quantum states from quantum noise by help of huge resource~\cite{Gottesman:1997ub,Gottesman:1998gt,Preskill:1998gf,Nielsen:2000vn,Paetznick:2013vs}.

We believe that a quantum computer component can work more efficiently and effectively if an input quantum state to the component is in a suited form for its function. 
For example, for a memory or a communication channel, more amount of information can be stored or transmitted if information is compressed.
On the other hand if the form of information is easy to handle fault tolerantly, then a central processing unit can manipulate it efficiently.
Here the form of information relies on a quantum error-correcting code.
A code rate $k/n$ and an ability to implement fault tolerant quantum gates of a quantum error-correcting code are closely associated with characteristics of an encoded quantum state.

A fault-tolerant quantum computing does not allow to decode an encoded quantum state and re-encode the decoded state with a different quantum error-correcting code in the middle of a computing because by doing so an unencoded quantum state is exposed to quantum noise without any shield.
By the way as mentioned before each quantum computer component prefers a quantum state in a certain form to others.
In this regards, Hill et al.~\cite{Hill:2013uj} proposed a fault-tolerant conversion between the $[[5,1,3]]$ stabilizer code and the $[[7,1,3]]$ Steane code. 
They mentioned that because a memory has a limited capacity using the $[[5,1,3]]$ code for a memory is more efficient while Steane code is more useful for manipulating a quantum state fault tolerantly in a central processing unit.
By applying a series of one- and two-qubit Clifford gates, the $[[5,1,3]]$ stabilizer code is transformed into the Steane code and vice versa.
By a Clifford operation, a quantum error-correcting code is converted to a Clifford equivalent code.
They argued that since all the \textit{possible} quantum errors at each step can be corrected, their conversion is fault tolerant.
Here the possible quantum errors mean an arbitrary one-qubit error and specific two-qubit error caused by a two-qubit Clifford gate.

On the other hand, Anderson et al.~\cite{Anderson:2014ja} employed a code conversion to implement a universal fault tolerant quantum computing.
Since there does not exist a quantum error-correcting code that can implement all universal quantum gates in a fault tolerant manner, they showed that by using a code conversion between two different quantum error-correcting codes, namely $[[15,1,3]]$ Reed-Muller code and Steane code, a set of universal quantum gates, $H$, $CNOT$ and $T$, can be implemented fault tolerantly.
Since Steane code belongs to the class of a quantum Reed-Muller code\footnote{By using the notation of the Ref.~\cite{Anderson:2014ja}, Steane code and $[[15,1,3]]$ Reed-Muller code respectively corresponds to QRM(3) and QRM(4).}, the conversion between both codes can be achieved by exploiting the structure of a Reed-Muller code.

In this work, we deal with a fault-tolerant conversion between stabilizer codes by applying Clifford gates.
Our conversion is motivated by Hill et al.~\cite{Hill:2013uj}, and our resulting circuit looks similar with theirs.
But, the principle to find a conversion circuit is completely different with their method.
Their algorithm first finds Clifford equivalent codes of the $[[5,1,3]]$ code and Steane code, and run an exhaustive computer search to find a conversion circuit.
The resulting circuit is optimal in terms of the number of two-qubit gates.
However, the applicability of the algorithm to other codes, in particular long codes, seems difficult because it is based on a computer search on Clifford equivalent codes.

On the other hand, our code conversion scheme is systematically based on the structure of a stabilizer code and Clifford gates.
The proposed scheme begins by putting stabilizer generators of both codes into specific forms.
The form are related to standard forms of the stabilizer codes.
The scheme then finds differences between both codes, and applies properly chosen Clifford gate.
A fault tolerance of the proposed scheme is verified by a successful quantum error correction in each step during the entire transformation.
A stabilizer code is transformed into a subsequent code by a Clifford gate and the subsequent code can correct all the possible quantum errors during the Clifford transformation.

The rest of this work is organized as follows.
Section~\ref{sec:stabilizer_formalism} reviews a stabilizer code and a Clifford group.
Section~\ref{sec:code_conversion} describes a code conversion scheme, and the fault tolerance of the scheme is discussed in Section~\ref{sec:fault_tolerance}.
Section~\ref{sec:case_study} shows three examples: the $[[5,1,3]]$ stabilizer code and $[[7,1,3]]$ Steane code, $[[7,1,3]]$ Steane code and $[[15,1,3]]$ Reed-Muller code, and $[[7,1,3]]$ Steane code and $(3,4)$-QPC code.
Section~\ref{sec:discussion} concludes this work with concluding remarks.

%
%
\section{Stabilizer code and Clifford group}\label{sec:stabilizer_formalism}

A pauli group over $n$ qubits  $\mathcal{P}_n$ is represented by a binary symplectic group without considering a phase factor~\cite{Gottesman:1997ub}.
Each one-qubit pauli element $I$, $X$, $Z$, and $Y$ is associated with a binary vector of length 2,
\begin{equation}\label{eq:single_qubit_pauli}
I=(0|0),~X=(1|0),~Z=(0|1),~\textrm{and}~Y=iXZ=(1|1).
\end{equation}
An $n$-qubit pauli element that is a tensor product of $n$ one-qubit pauli elements is associated with a binary vector of length $2n$.
A commutativity between pauli elements that two pauli elements are mutually commuting or anti-commuting is associated with the symplectic inner product between the corresponding binary symplectic vectors.
Given two binary vectors $\mathbf{u}=(\mathbf{a}|\mathbf{b})$ and $\mathbf{v}=(\mathbf{c}|\mathbf{d})$ where $\mathbf{a},~\mathbf{b},~\mathbf{c}$ and $\mathbf{d}$ $\in\mathbb{F}_2^{n}$, the symplectic inner product is defined as $\mathbf{u}\odot\mathbf{v}=\mathbf{a}\cdot\mathbf{d}+\mathbf{b}\cdot\mathbf{c}$ (mod 2) where $\mathbf{a}\cdot\mathbf{d}$ and $\mathbf{b}\cdot\mathbf{c}$ are the inner products of both vectors respectively.
If the product vanishes, $\mathbf{u}$ and $\mathbf{v}$ are mutually commuting.
Otherwise, they are anti-commuting.

Under the symplectic inner product condition, an $[[n, k, d]]$ stabilizer code that is defined by $n-k$ mutually commuting independent $n$-qubit pauli elements is described by a full rank binary matrix of size $(n-k)\times 2n$, $\mathcal{G}=(G_X|G_Z)$ where the sub-matrix $G_X~(G_Z)$ is associated with pauli-$X (Z)$. 
The commutativity condition of stabilizer generators related to $\mathcal{G}$ can be verified by showing $G_X\cdot {G_Z}^T+G_Z\cdot {G_X}^{T}=0$.

A Clifford group $\mathcal{C}$ is a normalizer of the pauli group $\mathcal{P}_n$ that transforms a pauli element to a pauli element under conjugation, $U\cdot p\cdot U^{\dagger}\in \mathcal{P}_n$ where $p\in \mathcal{P}_n$ and $U\in \mathcal{C}$~\cite{Gottesman:1997ub}.
In Table~\ref{tab:Cliffordgates}, we list the Clifford gates $H$, $P$, $CNOT$ and $CZ$. 
In particular, a $CZ$ gate that acts symmetrically over a control and a target qubit plays a significant role for the proposed conversion scheme.

\begin{table}[t]
\centering
\tcaption{One- and two-qubit Clifford gates. Note that $M$ is an arbitrary one-qubit pauli element, $M\in\{I, X, Z, Y\}$. In case of a two-qubit gate, we assume that the front qubit is a control qubit, and the latter qubit is a target qubit.}
\begin{tabular}{|c||c|c|} \hline 
Gate & Input & Output \\ \hline 
\multirow{3}{*}{$H$} & $X$ & $Z$ \\ \cline{2-3}
 & $Z$ & $X$ \\ \cline{2-3} 
 & $Y$ &$-Y$ \\ \cline{2-3} \hline
\multirow{3}{*}{$P$} & $X$ & $Y$ \\ \cline{2-3}
 & $Z$ & $Z$ \\ \cline{2-3}
 & $Y$ & $X$ \\ \cline{2-3} \hline
\multirow{2}{*}{$CNOT$} & $X\otimes M$ & $X\otimes XM$ \\ \cline{2-3} 
 & $M\otimes Z$ & $ZM\otimes Z$ \\ \cline{2-3} \hline
\multirow{2}{*}{$CZ$} & $X\otimes M$ & $X\otimes ZM$ \\ \cline{2-3}
 & $M\otimes X$ & $ZM\otimes X$ \\ \cline{2-3} \hline
\end{tabular}
\label{tab:Cliffordgates}
\end{table}

%
%

\section{Code conversion scheme}\label{sec:code_conversion}

An $[[n, k, d]]$ stabilizer code with stabilizer generators $\mathcal{G}=[G_X|G_Z]$ has a following standard form~\cite{Gottesman:1997ub,Nielsen:2000vn}
\begin{equation}\label{eq:standard_form}
\mathcal{G} = 
\left[
\begin{array}{ccc|ccc}
I & A_1 & A_2 & B & O & C \\
O & O & O & D & I & E
\end{array}
\right]
\begin{array}{l}
\}~r \\
\}~n-k-r
\end{array},
\end{equation}
where $r$ is the rank of $G_X$.
The logical $X$ and $Z$ operators of the code then are respectively defined as
\begin{equation}
\bar{X} = \left[
\begin{array}{ccc|ccc}
O & {E}^{T} & I & {C}^{T} & O & O
\end{array}
\right]
\end{equation}
and
\begin{equation}
\bar{Z} = \left[
\begin{array}{ccc|ccc}
O & O & O & {A_2}^{T} & O & I
\end{array}
\right].
\end{equation}
Given a stabilizer code, one can find its standard form by applying the gaussian elimination (including swapping qubits) to the stabilizer generators.
Note that an addition of row vectors during the gaussian elimination does not affect a stabilizer code space, but gives us another set of stabilizer generators.
On the other hand, swapping qubits permutes the code space.
Therefore, it should be involved in the code conversion circuit.

The standard form can be transformed into 
\begin{equation}
\left[
\begin{array}{ccc|ccc}
I & O & A_2 & B & A_1 & C \\
O & I & O & D & O & E
\end{array}
\right],
\end{equation}
by Hadamard gates on the qubits $r+1\sim n-k$.
Without loss of generality, we will denote the last matrix by 
\begin{equation}
\mathcal{G}_{IABC} = 
\left[
\begin{array}{cc|cc}
I & A & B & C
\end{array}
\right]
\begin{array}{r}
\}~n-k
\end{array}
\end{equation}
where $I$ is the identity matrix of size $(n-k)$ and $A$, $B$, $C$ are binary matrices of size $(n-k)\times k$, $(n-k)\times (n-k)$, and $(n-k)\times k$ respectively.
For the purpose of presentation, we call the stabilizer generators in the above form an \textit{IABC} form of a stabilizer code.
If the rank of $G_X$ is $n-k$, then the IABC form of the code is the same with the standard form.

Our code conversion scheme begins by putting both stabilizer codes into the IABC forms. 
Suppose that one transforms an $[[n_1, k_1, d_1]]$ stabilizer code $\mathcal{C}^{srcs}$ into an $[[n_2, k_2, d_2]]$ stabilizer code $\mathcal{C}^{trgt}$.
Note that to preserve the volume of information in the conversion, the dimensions of logical information of both codes have to be the same $k_1=k_2$.
By applying Clifford operations described above, one finds the IABC forms of both codes,
\begin{equation}\label{eq:IABC_both_codes}
\mathcal{G}^{srcs}_{IABC} = \left[
\begin{array}{cc|cc}
I & A_S & B_S & C_S
\end{array}
\right]
\end{equation}
and
\begin{equation}
\mathcal{G}^{trgt}_{IABC} = \left[
\begin{array}{cc|cc}
I & A_T & B_T & C_T
\end{array}
\right].
\end{equation}

Since both codes have different block size $n_1\neq n_2$, ancilla qubits have to be added to both codes 
\begin{equation}
\mathcal{C}^{srcs}\otimes |+\rangle^{\otimes m_1} \Leftrightarrow \mathcal{C}^{trgt}\otimes |+\rangle^{\otimes m_2},
\end{equation}
where $|+\rangle = (|0\rangle + |1\rangle) /\sqrt{2}$.
Therefore, the conversion scheme takes the augmented source code as a input code, and gives the augmented target code as an output.
After the whole conversion process, one can have a target code by deleting the last $m_2$ ancilla qubits.
The augmented codes are described by the following stabilizer generators,
\begin{equation}
\mathcal{G}^{srcs'} = \left[
\begin{array}{ccc|ccc}
I & A_S & O & B_S & C_S & O \\
O & O & I_{m_1} & O & O & O
\end{array}
\right]
\end{equation}
and 
\begin{equation}
\mathcal{G}^{trgt'} = \left[
\begin{array}{ccc|ccc}
I & A_T & O & B_T & C_T & O \\
O & O & I_{m_2} & O & O & O
\end{array}
\right].
\end{equation}

The IABC form of an augmented code is slightly different from unaugmented code.
By applying qubit swaps, CNOT and CZ gates, one finds the IABC form of the augmented code $\mathcal{C}^{srcs'}$
\begin{equation}
\mathcal{G}^{srcs'}_{IABC} = \left[
\begin{array}{ccc|ccc}
I & O & A_S & B_S & \Delta & C_S \\
O & I & \mathbb{1} & O & \Theta & \mathbb{1}
\end{array}
\right]~\begin{array}{l}
\}~n_1-k \\
\}~m_1
\end{array}
,
\end{equation}
where $\mathbb{1}$ is a matrix all the elements of which are 1.
The matrices $\Delta$ and $\Theta$ are raised by the CNOT and CZ gates from the zero matrices.
This matrix is also in the form IABC,
\begin{equation}
\mathcal{G}^{srcs'}_{IABC} = \left[
\begin{array}{cc|cc}
I & {A_S}' & {B_S}' & {C_S}'
\end{array}
\right].
\end{equation}
The IABC form of the augmented target code similarly can be derived
\begin{equation}
\mathcal{G}^{trgt'}_{IABC} = \left[
\begin{array}{cc|cc}
I & {A_T}' & {B_T}' & {C_T}'
\end{array}
\right].
\end{equation}
The number of ancilla qubits relies on the codes.
To conclude, an augmented code can be put in the IABC form by applying a Clifford operation $\mathcal{U}$ that consists of several Clifford gates,
\begin{equation}
\mathcal{G}^{srcs'}_{IABC} = \mathcal{U}^{srcs}\cdot \Bigl(\mathcal{C}^{srcs}\otimes |+\rangle^{\otimes m_1}\Bigr),
\end{equation}
and
\begin{equation}
\mathcal{G}^{trgt'}_{IABC} = \mathcal{U}^{trgt}\cdot \Bigl(\mathcal{C}^{trgt}\otimes |+\rangle^{\otimes m_2}\Bigr).
\end{equation}

The next step of the conversion finds the differences between $\mathcal{G}^{srcs'}_{IABC}$ and $\mathcal{G}^{trgt'}_{IABC}$, and applies Clifford gates according to the differences.
The applications of CNOT gates by the difference between $A'_S$ and $A'_T$, and CZ gates by the difference between $C'_S$ and $C'_T$ transforms $\mathcal{G}^{srcs'}_{IABC}$ into
\begin{equation}\label{eq:transformed_source_code}
\mathcal{G}^{srcs''}_{IABC} = 
\left[
\begin{array}{cc|cc}
I & A'_T & B''_S & C'_T
\end{array}\right].
\end{equation}
The matrix $B''_S$ is raised by applications of the CNOT and CZ gates from $B'_S$.
The IABC form of the augmented source code has been transformed into that of the augmented target code except the matrix $B''_S$.
The remaining difference between $B''_S$ and $B'_T$ then can be resolved by CZ operations if a difference matrix $D$, defined as $B''_S+B'_T$ (mod 2), is symmetric.
If $D$ is symmetric, 
\begin{equation}
\mathcal{G}^{trgt'}_{IABC} = CZ(q_1, q_2)\cdot \mathcal{G}^{srcs''}_{IABC}
\end{equation}
over qubits $(q_1, q_2)$ where $D[q_1,q_2]=1$ for $q_1<q_2$.

Then, what is a condition for the difference matrix $D$ to be symmetric?
From the symplectic inner product condition over the IABC forms, one reads that
\begin{equation}
(I~A'_T) \times {(B''_S~C'_T)}^{T} + (B''_S~C'_T)\times {(I~A'_T)}^{T} = {B''_S}^{T} + A'_T {C'_T}^{T} + B''_S + C'_T {A'_T}^{T} = 0
\end{equation}
from $\mathcal{G}^{srcs''}_{IABC}$, and similarly
\begin{equation}
{B'_T}^{T} + A'_T {C'_T}^{T} + B'_T + C'_T {A'_T}^{T} = 0
\end{equation}
from $\mathcal{G}^{trgt'}_{IABC}$.
Then, one knows that the difference matrix $D$ is always symmetric by adding both equations in mod 2,
\begin{equation}
{B''_S}^{T} +B''_S + {B'_T}^{T} + B'_T = (B''_S + B'_T)^{T} + (B''_S + B'_T) = 0.
\end{equation}
Therefore the present scheme is applicable to a pair of any stabilizer codes in principle.

Now, the IABC form of the augmented source code has been completely transformed into that of the augmented target code.
But, it is not enough because the IABC form is not the exact stabilizer generators of the target code.
For the complete conversion, an additional application of the inversion of the Clifford operations used to find the IABC form of the augmented target code are required,
\begin{equation}
\mathcal{G}^{trgt}\otimes |+\rangle^{\otimes m_2} = {\mathcal{U}^{trgt}}^{\dag}\cdot \mathcal{G}^{trgt'}_{IABC}.
\end{equation}

%
%

\section{On the fault-tolerance of the proposed scheme}\label{sec:fault_tolerance}

The fault tolerance of the proposed scheme can be guaranteed by a quantum error correction in each step during the entire conversion.
In each step, by a Clifford gate a stabilizer code is transformed into a Clifford equivalent stabilizer code.
For the fault tolerance, the following code has to correct all the possible quantum errors happened during the Clifford operation.
What we have to consider particularly is that the present code conversion scheme involves many two-qubit Clifford gates CNOT and CZ.
As shown in Table~\ref{tab:Cliffordgates}, such a two-qubit gate transforms an one-qubit error to a two-qubit error if the one-qubit error happens on a qubit where the two-qubit gate is acting.
To conclude, a stabilizer code obtained by an one-qubit Clifford gate has to correct an arbitrary one-qubit error, but a stabilizer code by a two-qubit Clifford gate has to correct some specific two-qubit error caused by the gate in addition to an arbitrary one-qubit error.

For the quantum error correction, we need to discuss two conditions.
First, all the stabilizer codes in the middle of the conversion have to have minimum distance at least 3 to correct an arbitrary one-qubit error and specific two-qubit errors caused by a two-qubit Clifford gate.
In general a stabilizer code of minimum distance 5 is required to correct an arbitrary two-qubit quantum errors.
But, in the middle of the proposed code conversion, only some specific two-qubit errors happen by a two-qubit Clifford gate.
The two-qubit error can be exactly corrected if it has a unique error syndrome or \textit{degenerate} with an one-qubit error.
If quantum errors $E$ and $F$ are related by a stabilizer operator $S$ as $F=E\cdot S$, it is called that both errors are degenerate.
Two degenerate quantum errors act on a code space in the same way, and both can be corrected by one recovery,
\begin{equation}
F|\psi\rangle = E\cdot S|\psi\rangle \leftrightarrow F|\psi\rangle = E|\psi\rangle.
\end{equation}

Second, all the possible quantum errors must not belong to $N(\mathcal{S})-\mathcal{S}$ where $\mathcal{S}$ is the stabilizer and $N(\mathcal{S})$ is the normalizer of $\mathcal{S}$.
A quantum error that belongs to the normalizer but outside of the stabilizer is not detected by a syndrome measurement by definition.
But, it modifies the code space.
Therefore, such an error which corresponds to a logical operator of the code should not happen during the code conversion.

In the previous section, we described that two-qubit Clifford operations transform the sub-matrices of $\mathcal{C}^{srcs'}_{IABC}$ to the corresponding sub-matrices of $\mathcal{C}^{trgt'}_{IABC}$.
There we did not mention anything about the order of the operations.
By the way the order is very important for the fault tolerance of the code conversion.
It happens that a certain two-qubit operation makes a following code has a minimum distance less than 3 or a CZ operation flips all $Z$s on a qubit to all $I$s.
Fortunately, reordering the gates properly resolves the problems, and makes a code conversion circuit fault tolerant.
In the following section, we will show fault tolerant circuits for the examples.

%
%

\section{Case studies}\label{sec:case_study}


\subsection{$[[5,1,3]]$ stabilizer code and $[[7,1,3]]$ Steane code}\label{subsec:steane_5qubit}

In this section, we show the $[[5,1,3]]$ stabilizer code $\mathcal{C}^{[[5,1,3]]}$ is transformed into $[[7, 1, 3]]$ Steane code $\mathcal{C}^{Steane}$.
This conversion was also covered in Ref.~\cite{Hill:2013uj}.
The $[[5,1,3]]$ code is usually represented by the following stabilizer generators in literature~\cite{Gottesman:1997ub,Nielsen:2000vn,Lidar:2013ts},
\begin{eqnarray} 
\mathcal{G}^{[[5,1,3]]} &=& 
\left(
\begin{array}{ccccc}
X & Z & Z & X & I \\
I & X & Z & Z & X \\
X & I & X & Z & Z \\
Z & X & I & X & Z
\end{array}
\right) \\
&=&	
\left[
\begin{array}{ccccc|ccccc}
1 & 0 & 0 & 1 & 0 & 0 & 1 & 1 & 0 & 0\\
0 & 1 & 0 & 0 & 1 & 0 & 0 & 1 & 1 & 0 \\
1 & 0 & 1 & 0 & 0 & 0 & 0 & 0 & 1 & 1 \\
0 & 1 & 0 & 1 & 0 & 1 & 0 & 0 & 0 & 1
\end{array}
\right].
\end{eqnarray}
The stabilizer generators can be transformed into the standard form
\begin{equation}
\mathcal{G}^{[[5,1,3]]}= 
\left[
\begin{array}{ccccc|ccccc}
1 & 0 & 0 & 0 & 1 & 1 & 1 & 0 & 1 & 1 \\
0 & 1 & 0 & 0 & 1 & 0 & 0 & 1 & 1 & 0 \\
0 & 0 & 1 & 0 & 1 & 1 & 1 & 0 & 0 & 0 \\
0 & 0 & 0 & 1 & 1 & 1 & 0 & 1 & 1 & 1
\end{array}
\right],
\end{equation}
without any swap operation. 
As mentioned before, the IABC form of the $[[5,1,3]]$ code is the same with the standard form because the rank of $G_X$ equals to the number of stabilizer generators.

For the conversion from the $[[5,1,3]]$ code to Steane code three ancilla qubits are added to the $[[5,1,3]]$ code, and one ancilla qubit is necessary for Steane code,
\begin{equation}
\mathcal{C}^{[[5,1,3]]}\otimes |+\rangle^{\otimes 3} \Leftrightarrow \mathcal{C}^{Steane}\otimes |+\rangle.
\end{equation}
The augmented $[[5,1,3]]$ code is defined by the following stabilizer generators, 
\begin{equation} 
\mathcal{G}^{[[5,1,3]]'}= 
\left[
\begin{array}{ccccc|ccc|ccccc|ccc}
1 & 0 & 0 & 0 & 1 & 0 & 0 & 0 &	1 & 1 & 0 & 1 & 1 & 0 & 0 & 0\\
0 & 1 & 0 & 0 & 1 & 0 & 0 & 0 &	0 & 0 & 1 & 1 & 0 & 0 & 0 & 0\\
0 & 0 & 1 & 0 & 1 & 0 & 0 & 0 &	1 & 1 & 0 & 0 & 0 & 0 & 0 & 0\\
0 & 0 & 0 & 1 & 1 & 0 & 0 & 0 &	1 & 0 & 1 & 1 & 1 & 0 & 0 & 0\\ \hline
0 & 0 & 0 & 0 & 0 & 1 & 0	& 0 & 0 & 0 & 0 & 0 & 0 & 0 & 0 & 0\\
0 & 0 & 0 & 0 & 0 & 0 & 1	& 0 & 0 & 0 & 0 & 0 & 0 & 0 & 0 & 0\\
0 & 0 & 0 & 0 & 0 & 0 & 0	& 1 & 0 & 0 & 0 & 0 & 0 & 0 & 0 & 0\\
\end{array}
\right].
\end{equation}
Three swap operations, SWAP(5, 6), SWAP(6, 7) and SWAP(7, 8), transform the above generators into the following form,
\begin{eqnarray} 
\mathcal{G}^{[[5,1,3]]'}= 
\left[
\begin{array}{cccccccc|cccccccc}
1 & 0 & 0 & 0 & 0 & 0 & 0 & 1 &	1 & 1 & 0 & 1 & 0 & 0 & 0 & 1\\
0 & 1 & 0 & 0 & 0 & 0 & 0 & 1 &	0 & 0 & 1 & 1 & 0 & 0 & 0 & 0 \\
0 & 0 & 1 & 0 & 0 & 0 & 0 & 1 &	1 & 1 & 0 & 0 & 0 & 0 & 0 & 0\\
0 & 0 & 0 & 1 & 0 & 0 & 0 & 1 &	1 & 0 & 1 & 1 & 0 & 0 & 0 & 1 \\
0 & 0 & 0 & 0 & 1 & 0 & 0 & 0 & 0 & 0 & 0 & 0 & 0 & 0 & 0 & 0 \\
0 & 0 & 0 & 0 & 0 & 1 & 0 & 0 & 0 & 0 & 0 & 0 & 0 & 0 & 0 & 0 \\
0 & 0 & 0 & 0 & 0 & 0 & 1 & 0 & 0 & 0 & 0 & 0 & 0 & 0 & 0 & 0 \\
\end{array}
\right].
\end{eqnarray}
Additionally, CNOT and CZ operations to the 8-th qubit conditioned on the $5\sim 7$-qubits introduce the following IABC form of the augmented $[[5,1,3]]$ code
\begin{eqnarray}
\mathcal{G}^{[[5,1,3]]'}_{IABC} &=& [I~A'_5|B'_5~C'_5] \\
&=& 
\left[
\begin{array}{cccccccc|cccccccc}
1 & 0 & 0 & 0 & 0 & 0 & 0 & 1 &		1 & 1 & 0 & 1 & 0 & 0 & 0 & 1 \\
0 & 1 & 0 & 0 & 0 & 0 & 0 & 1 &		0 & 0 & 1 & 1 & 1 & 1 & 1 & 0 \\
0 & 0 & 1 & 0 & 0 & 0 & 0 & 1 &		1 & 1 & 0 & 0 & 1 & 1 & 1 & 0 \\
0 & 0 & 0 & 1 & 0 & 0 & 0 & 1 &		1 & 0 & 1 & 1 & 0 & 0 & 0 & 1 \\
0 & 0 & 0 & 0 & 1 & 0 & 0 & 1 & 	0 & 0 & 0 & 0 & 1 & 1 & 1 & 1 \\
0 & 0 & 0 & 0 & 0 & 1 & 0 & 1 & 	0 & 0 & 0 & 0 & 1 & 1 & 1 & 1 \\
0 & 0 & 0 & 0 & 0 & 0 & 1 & 1 & 	0 & 0 & 0 & 0 & 1 & 1 & 1 & 1 \\
\end{array}
\right].
\end{eqnarray}

To find the IABC form of the augmented Steane code, the following Clifford gates are required.
First, two swap operations, SWAP(4,5) and SWAP(6,7), are used to find the standard form during the gaussian elimination.
Then, the application of SWAP(7,8), CNOT(7,8) and CZ(7,8) transforms the standard form into the IABC form of the augmented Steane code,
\begin{eqnarray}\label{eq:IABC_steane} 
\mathcal{G}^{Steane'}_{IABC} &=& [I~A'_{St}|B'_{St}~C'_{St}] \\
&=& 
\left[
\begin{array}{cccccccc|cccccccc}
1 & 0 & 0 & 0 & 0 & 0 & 0 & 0 	& 0 & 0 & 0 & 1 & 1 & 1 & 0 & 0\\ 
0 & 1 & 0 & 0 & 0 & 0 & 0 & 1 	& 0 & 0 & 0 & 1 & 0 & 1 & 1 & 0\\ 
0 & 0 & 1 & 0 & 0 & 0 & 0 & 1 	& 0 & 0 & 0 & 0 & 1 & 1 & 1 & 0\\ 
0 & 0 & 0 & 1 & 0 & 0 & 0 & 0 	& 1 & 0 & 1 & 0 & 0 & 0 & 1 & 1\\
0 & 0 & 0 & 0 & 1 & 0 & 0 & 0 	& 1 & 1 & 0 & 0 & 0 & 0 & 1 & 1\\
0 & 0 & 0 & 0 & 0 & 1 & 0 & 0 	& 1 & 1 & 1 & 0 & 0 & 0 & 0 & 0\\ 
0 & 0 & 0 & 0 & 0 & 0 & 1 & 1 	& 0 & 0 & 0 & 0 & 0 & 0 & 1 & 1\\ 
\end{array}
\right].
\end{eqnarray}

Now it is time to transform the IABC form of the augmented $[[5,1,3]]$ code to that of the augmented Steane code.
The IABC form the augmented $[[5,1,3]]$ code is transformed into
\begin{eqnarray} 
\mathcal{G}^{[[5,1,3]]''}_{IABC} &=& [I~A'_{St}|B''_5~C'_{St}] \\
&=&	
\left[
\begin{array}{cccccccc|cccccccc}
1 & 0 & 0 & 0 & 0 & 0 & 0 & 0 &		0 & 1 & 0 & 0 & 1 & 1 & 0 & 0 \\
0 & 1 & 0 & 0 & 0 & 0 & 0 & 1 &		1 & 0 & 1 & 1 & 1 & 0 & 1 & 0 \\
0 & 0 & 1 & 0 & 0 & 0 & 0 & 1 &		0 & 1 & 0 & 0 & 1 & 0 & 1 & 0 \\
0 & 0 & 0 & 1 & 0 & 0 & 0 & 0 &		0 & 0 & 1 & 0 & 1 & 1 & 0 & 1 \\
0 & 0 & 0 & 0 & 1 & 0 & 0 & 0 & 	1 & 0 & 0 & 1 & 0 & 0 & 1 & 1 \\
0 & 0 & 0 & 0 & 0 & 1 & 0 & 0 & 	1 & 0 & 0 & 1 & 0 & 0 & 1 & 0 \\
0 & 0 & 0 & 0 & 0 & 0 & 1 & 1 & 	0 & 0 & 0 & 1 & 0 & 1 & 1 & 1 \\
\end{array}
\right].
\end{eqnarray}
by CNOT(5,8), CNOT(4,8), CNOT(6,8), CNOT(1,8), CZ(1,8) and CZ(6,8) in the named order.
It is almost the same with $\mathcal{G}^{Steane'}_{IABC}$ except the matrix $B'_{St}$.
As mentioned before, CZ operations can resolve the difference because the difference matrix $D=B''_5+B'_{St}$ is symmetric as follows,
\begin{equation} 
D=\left[
\begin{array}{ccccccc}
0 & 1&  0&  1&  0&  0&  0\\
1 & 0 & 1 & 0 & 1 & 1 & 0 \\
0 & 1 & 0 & 0 & 0 & 1 & 0 \\
1 & 0 & 0 & 0 & 1 & 1 & 1 \\
0 & 1 & 0 & 1 & 0 & 0 & 0 \\
0 & 1 & 1 & 1&  0&  0&  1 \\
0 & 0 & 0 & 1 & 0 & 1 & 0
\end{array}
\right].
\end{equation}
The required CZ gates are CZ(1,4), CZ(2,5), CZ(1,2), CZ(3,6), CZ(4,5), CZ(4,6), CZ(4,7), CZ(2,6), CZ(6,7) and CZ(2,3) in the named order.

\begin{figure}[t]
\centering
\epsfig{file=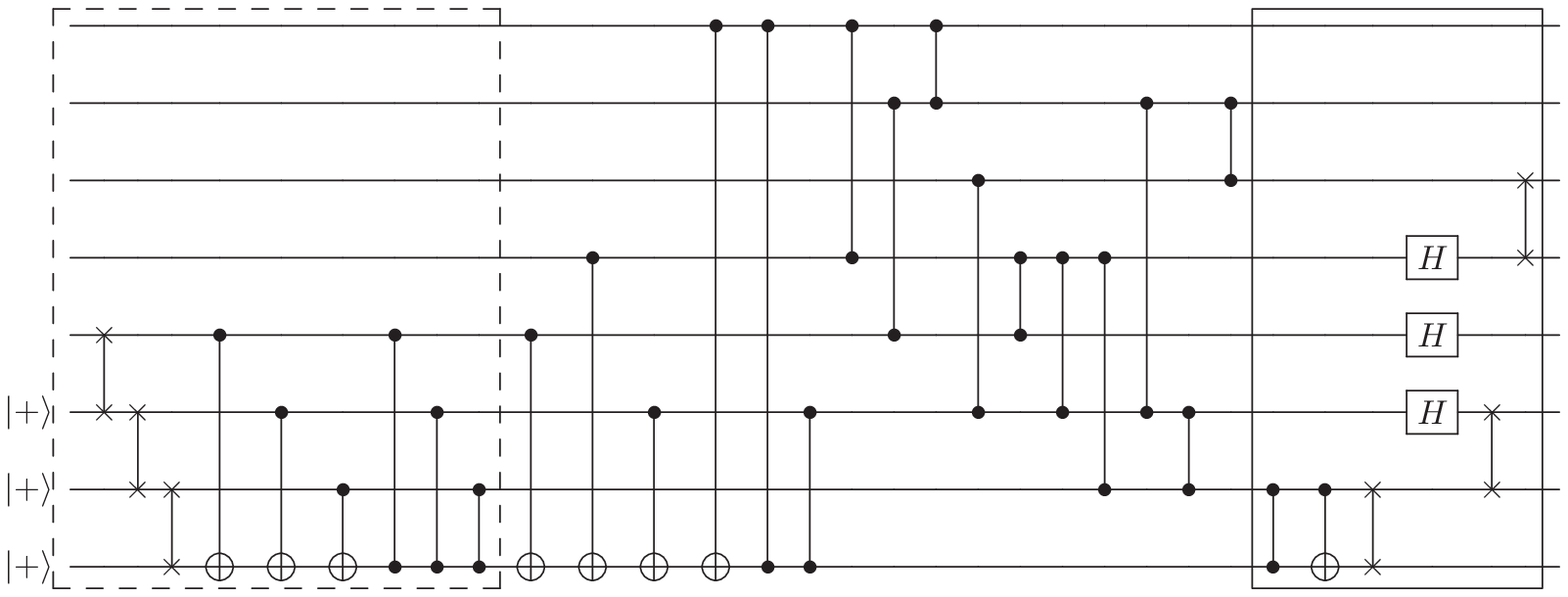, scale=0.8}
\fcaption{A fault-tolerant quantum circuit to transform the $[[5,1,3]]$ code to Steane code. 
The input state at the most left side is a tensor product of a logical qubit of the $[[5,1,3]]$ code and ancilla qubits $|+\rangle^{\otimes 3}$, and the resulting state at the most right side is a logical qubit of Steane code.
The box of the dotted line indicates Clifford operations to find the IABC form of the augmented $[[5,1,3]]$ code, and that of the solid line represents the inverse of the Clifford operations to find the IABC form of the augmented Steane code.}
\label{fig:5qubit_to_steane}
\end{figure}

The IABC form of the augmented $[[5,1,3]]$ code has been transformed into that of the augmented Steane code now.
As a final step, the application of the inverse operation of the Clifford operations used to find $\mathcal{C}^{Steane'}_{IABC}$ is required.
The Clifford operations are SWAP(3,4), SWAP(6,7), H(4), H(5), H(6), SWAP(7,8), CNOT(7,8) and CZ(7,8). 
Applying these gates in the reverse order completes the entire conversion.

Fig.~\ref{fig:5qubit_to_steane} shows a fault tolerant circuit that transforms the $[[5,1,3]]$ code into Steane code.
The box of the dotted line in the figure indicates the Clifford gates to find the IABC form of the augmented $[[5,1,3]]$ code, and the box of the solid line represents the inverse operation of the Clifford gates to find the IABC form of the augmented Steane code.
We summarize the Clifford gates for this conversion in Table~\ref{tab:clifford_list_5qubit_to_steane}.
The conversion of the opposite direction can be achieved by applying the circuit in the reverse order.

Let us check the fault tolerance of the code conversion.
As an example, the code obtained by CZ(6,8), in the IABC form, is defined by the following stabilizer generators
\begin{equation}
\begin{array}{cccccccc}
X& Z& I& I& Z& Z& I& I \\
Z& X& Z& Z& Z& I& Z& X \\
I& Z& X& I& Z& I& Z& X \\
I& I& Z& X& Z& Z& I& Z \\
Z& I& I& Z& X& I& Z& Z \\
Z& I& I& Z& I& X& Z& I \\
I& I& I& Z& I& Z& Y& Y
\end{array}
\end{equation}
and its logical Pauli operators are
\begin{equation}
\begin{array}{ccccccccc}
\bar{X} = & I& I& I& Z& Z& I& Z& X\\
\bar{Z} = & I& Z& Z& I& I& I& Z& Z
\end{array}.
\end{equation}
Since the minimum distance of this code is 3, an arbitrary one-qubit error is successfully corrected.
By CZ(6,8), possible two-qubit errors are $X_6 Z_8$, $Z_6 X_8$, $Y_6 Z_8$ and $Z_6 Y_8$.
$X_6 Z_8$ is degenerate with the one-qubit error $X_5$ with a common error syndrome $(-1, -1, -1, -1, +1, +1, +1)$, and the other two qubit errors have its own unique error syndrome.
To conclude, the code by CZ(6,8) can correct all the possible quantum errors.

\begin{table}[t]
\centering
\tcaption{A list of the Clifford gates for the conversion from the $[[5,1,3]]$ code to Steane code. All the gates have to be applied in the written order. }
\begin{tabular}{|c||c|} \hline 
Purpose & Gates \\ \hline 
\multirow{3}{*}{$\mathcal{G}^{[[5,1,3]]'}_{IABC}$} & SWAP(5,6), SWAP(6,7), SWAP(7,8), \\ 
									  & CNOT(5,8), CNOT(6,8), CNOT(7,8), \\ 
									  & CZ(5,8), CZ(6,8), CZ(7,8) \\ \hline

\multirow{2}{*}{$\mathcal{G}^{Steane}_{IABC}$} & SWAP(3,4), SWAP(6,7), H(4), H(5), H(6), \\ 
									 & SWAP(7,8), CNOT(7,8), CZ(7,8) \\  \hline

$A'_5\rightarrow A'_{St}$ & CNOT(5,8), CNOT(4,8), CNOT(6,8), CNOT(1,8) \\ \hline

$C'_5\rightarrow C'_{St}$ & CZ(1,8), CZ(6,8)\\ \hline

\multirow{2}{*}{$B''_5 \rightarrow B'_{St}$} & CZ(1,4), CZ(2,5), CZ(1,2), CZ(3,6), CZ(4,5), \\ 
								& CZ(4,6), CZ(4,7), CZ(2,6), CZ(6,7), CZ(2,3) \\ \hline

\end{tabular}
\label{tab:clifford_list_5qubit_to_steane}
\end{table}

\begin{figure}[t]
\centering
\epsfig{file=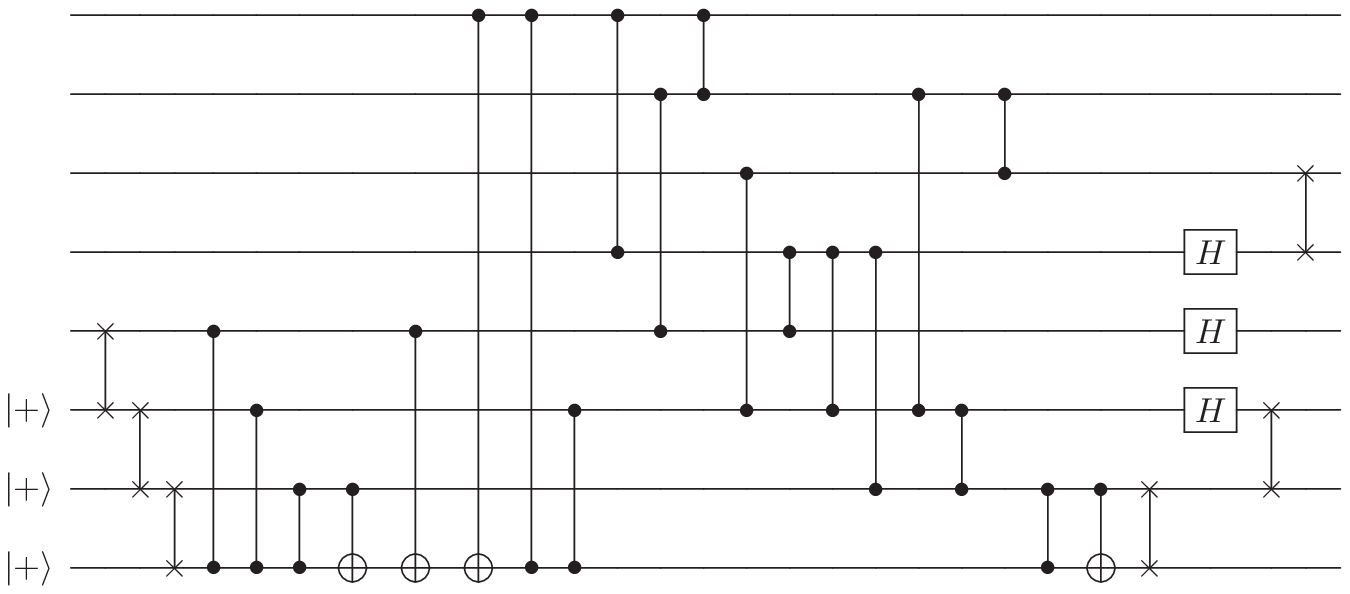, scale=0.8}
\fcaption{A simplified fault-tolerant quantum circuit to transform the $[[5,1,3]]$ code to Steane code. Four two-qubit gates are removed from the circuit of Fig.~\ref{fig:5qubit_to_steane}.}
\label{fig:5qubit_to_steane_opt}
\end{figure}

The proposed scheme is systematic based on the IABC forms of the codes.
By a routine of the scheme, it happens that a gate is required several times.
For example, CNOT(5,8) and CNOT(6,8) are involved twice in the circuit of Fig.~\ref{fig:5qubit_to_steane}.
If the position of those gates can be re-arranged fault tolerantly, the effect of both gates can be canceled out and therefore can be removed from a circuit.
Fortunately, in the box of the dotted line of the figure, CNOT gates can be placed after CZ gates, and the second CNOT(6,8) can be placed in front of CNOT(4,8).
Therefore, two repetitive CNOT(5,8)s can CNOT(6,8)s do not affect the code conversion, and therefore can be removed.
The simplified circuit is shown in Fig.~\ref{fig:5qubit_to_steane_opt}.

\subsection{$[[7,1,3]]$ Steane code and $[[15, 1 ,3]]$ Reed-Muller code}\label{sec:steane_rm}

The conversion between Steane code and $[[15,1,3]]$ Reed-Muller code was covered in Ref.~\cite{Anderson:2014ja} to implement a universal fault tolerant quantum computing.
It is well known that Steane code belongs to the class of a quantum Reed-Muller code, and $[[15,1,3]]$ Reed-Muller code can be defined by Steane code.
Anderson et al.~\cite{Anderson:2014ja} exploited the code structure of quantum Reed-Muller codes for the conversion.
To be exact, a specially designed ancilla state $|\Phi\rangle=1/\sqrt{2}(|\bar{0}\rangle_m |0\rangle + |\bar{1}\rangle_m|1\rangle)$ plays a significant role where $|\bar{i}\rangle_m$ is a logical qubit $i$.
By attaching the ancilla state, the augmented Steane code becomes so-called an ``extended quantum Reed-Muller code".
This code can be transformed into the target Reed-Muller code by changing stabilizer generators.
This scheme does not require Clifford gates for the code conversion, but the preparation of the ancilla states $|\Phi\rangle$ is non-negligible.

\begin{figure}[t]
\centering
\epsfig{file=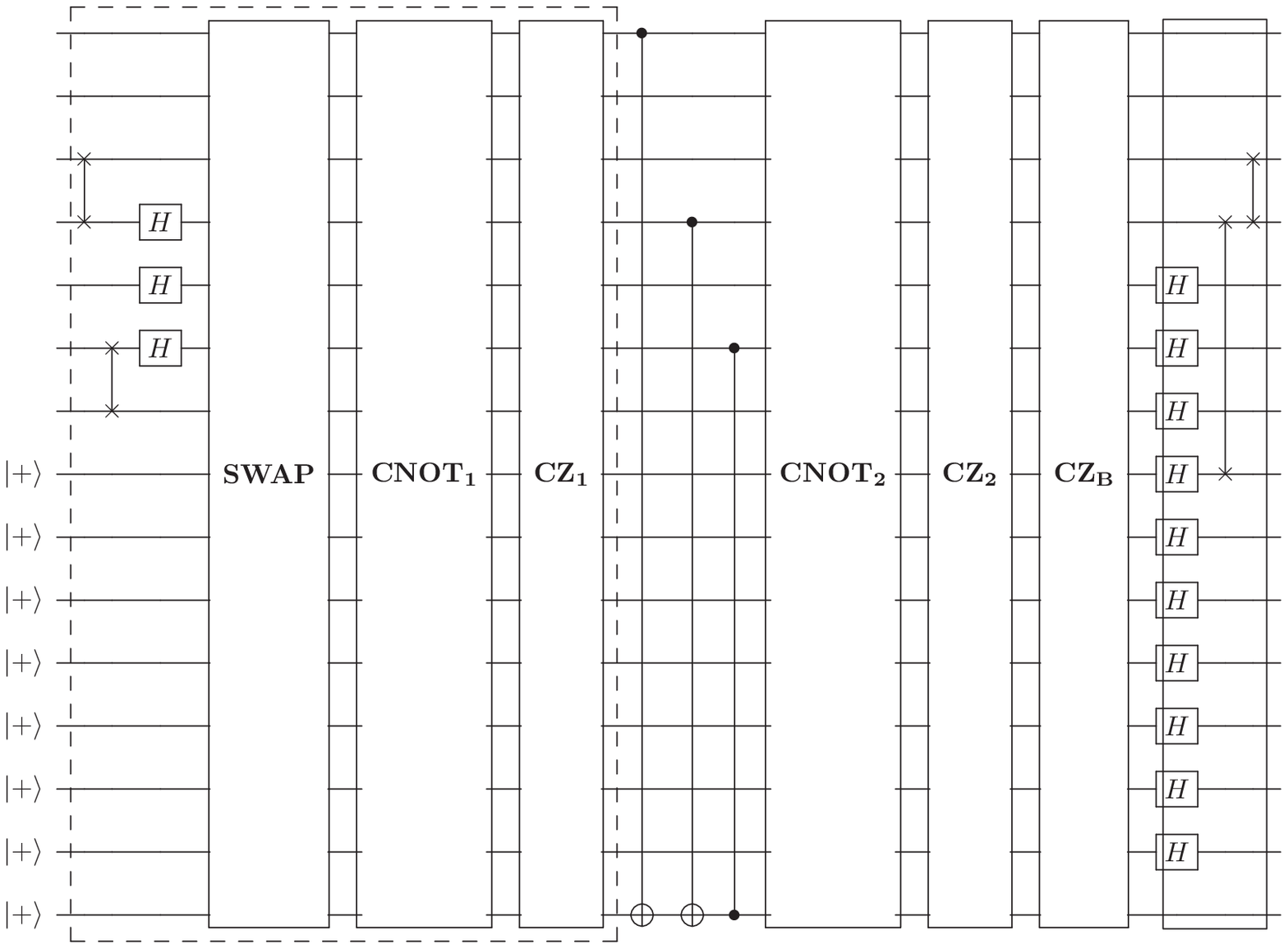,scale=0.8}
\fcaption{A quantum circuit to convert Steane code to $[[15,1,3]]$ Reed-Muller code.
The input state at the left side is a tensor product of a logical qubit of Steane code and $|+\rangle^{\otimes 8}$, and the output state at the ther side is a logical qubit of the Reed-Muller code. 
The box of the dotted line indicates Clifford gates to find the IABC form of the augmented Steane code, and the box of the solid line represents the inverse of the Clifford gates used to find the IABC form of $[[15,1,3]]$ Reed-Muller code. The details of $\mathbf{SWAP}$, $\mathbf{CNOT_1}$, $\mathbf{CZ_1}$, $\mathbf{CNOT_2}$, $\mathbf{CZ_2}$ and $\mathbf{CZ_B}$ can be found in Tab.~\ref{tab:clifford_list_steane_to_rm}.}
\label{fig:steane_to_rm}
\end{figure}

To transform Steane code into $[[15,1,3]]$ Reed-Muller code by using the proposed scheme, one first has to add $m_1=8$ ancilla qubits to Steane code.
\begin{equation}
\mathcal{C}^{Steane}\otimes |+\rangle^{\otimes 8} \Leftrightarrow \mathcal{C}^{RM}.
\end{equation}
$[[15,1,3]]$ Reed-Muller code is a Calderbank-Shor-Steane (CSS) code defined by the following stabilizer generators
\begin{equation}
\mathcal{G}^{RM}=\left[
\begin{array}{c|c}
\bar{G}_X & 0 \\
0 & \bar{G}_Z
\end{array}
\right],
\end{equation}
where $G_X$ is a generator matrix of the classical Reed-Muller code RM(1,4), and $\bar{G}_X$ is a punctured generator matrix by removing the first row and column of $G_X$\cite{Anderson:2014ja}.
$G_Z$ is a generator matrix of RM(2,4) that is a dual code of RM(1,4), and $\bar{G}_Z$ is similarly defined.

Without details during the conversion, we list the required Clifford gates in Tab.~\ref{tab:clifford_list_steane_to_rm}.
The conversion from Steane code to $[[15,1,3]]$ Reed-Muller code completes by applying the Clifford gates in the named order.
Fig.~\ref{fig:steane_to_rm} shows the corresponding circuit.
The reverse conversion can be achieved by applying the gates in the reverse order.

\begin{sidewaystable}[t]
\centering
\tcaption{The list of the Clifford gates for the conversion from Steane code to $[[15,1,3]]$ Reed-Muller code. All the gates have to be applied in the named order. 
In this conversion, the Clifford gates for the conversions $A'_{St}\rightarrow A'_{RM}$ and $C'_{St}\rightarrow C'_{RM}$ are not separately applied.}
\begin{tabular}{|c||c|} \hline 
Purpose & Gates \\ \hline 
\multirow{2}{*}{$\mathcal{G}^{Steane'}_{IABC}$} & SWAP(3,4), SWAP(6,7), H(4), H(5), H(6), $\mathbf{SWAP}$=\{SWAP($i,i+1$) for $i=7\cdots14$\}, \\ 
& $\mathbf{CNOT_1}$=\{CNOT($i,15$) for $i=7\sim 14$\}, $\mathbf{CZ_1}$=\{CZ($i,15$) for $i=7\sim 14$\} \\ \hline

$\mathcal{G}^{RM}_{IABC}$ &  SWAP(3,4), SWAP(4,8), \{H($i$) for $i=5\cdots14$\} \\  \hline

\multirow{3}{*}{$A'$, $C'$}  & CNOT(1,15), CNOT(4,15), CZ(6,15),~$\mathbf{CNOT_2}$=\{CNOT($i,15$) for $i=7\sim 14$\},\\ 
 & $\mathbf{CZ_2}$=\{CZ(4,15), CZ(7,15), CZ(11,15), CZ(13, 15), CZ(14,15) \} \\ \hline

\multirow{3}{*}{$B''_{St} \rightarrow B_{RM}$} 	& $\mathbf{CZ_B}$=\{CZ(1,8), CZ(1,9), CZ(2,7), CZ(2,6), CZ(2,9), CZ(2, 11), CZ(2,12), \\ 
									& CZ(2, 14), CZ(1, 14), CZ(3,7), CZ(3,4), CZ(3,8), CZ(3,6), CZ(3,9), CZ(3,10), \\ 
									& CZ(3,13), CZ(3,14), CZ(4,6), CZ(4,7), CZ(4,9), CZ(4,10), CZ(4,12), CZ(6,7),  \\
									& CZ(6,8), CZ(6,9), CZ(6,10), CZ(6,11), CZ(6,12), CZ(6,13), CZ(6,14)\}\\ \hline
\end{tabular}
\label{tab:clifford_list_steane_to_rm}
\end{sidewaystable}


\subsection{$[[7,1,3]]$ Steane code and $(3,4)$-QPC code}\label{sec:steane_shor}

An $(n,m)$ quantum parity check (QPC) code~\cite{Knill:2001is,Ralph:2005ie,Muralidharan:2014gk} is a quantum error-correcting code with the following encoded qubits
\begin{equation}
|0\rangle_L = \frac{1}{\sqrt{2}}\Bigl(|+\rangle_L + |-\rangle_L \Bigr),~
|1\rangle_L = \frac{1}{\sqrt{2}}\Bigl(|+\rangle_L - |-\rangle_L \Bigr),
\end{equation}
where 
\begin{eqnarray}
|+\rangle_L &=& \frac{1}{\sqrt{2^n}} \Bigl(|0\rangle^{\otimes m} + |1\rangle^{\otimes m} \Bigr)^{\otimes n} \\
|-\rangle_L &=& \frac{1}{\sqrt{2^n}} \Bigl(|0\rangle^{\otimes m} - |1\rangle^{\otimes m} \Bigr)^{\otimes n}. 
\end{eqnarray}
The QPC code has been considered to correct photon losses~\cite{Knill:2001is,Ralph:2005ie,Kok:2007ep,Munro:2012gu,Muralidharan:2014gk}, and furthermore to implement a fault tolerant quantum repeater~\cite{Muralidharan:2014gk}.

\begin{figure}[t]
\centering
\epsfig{file=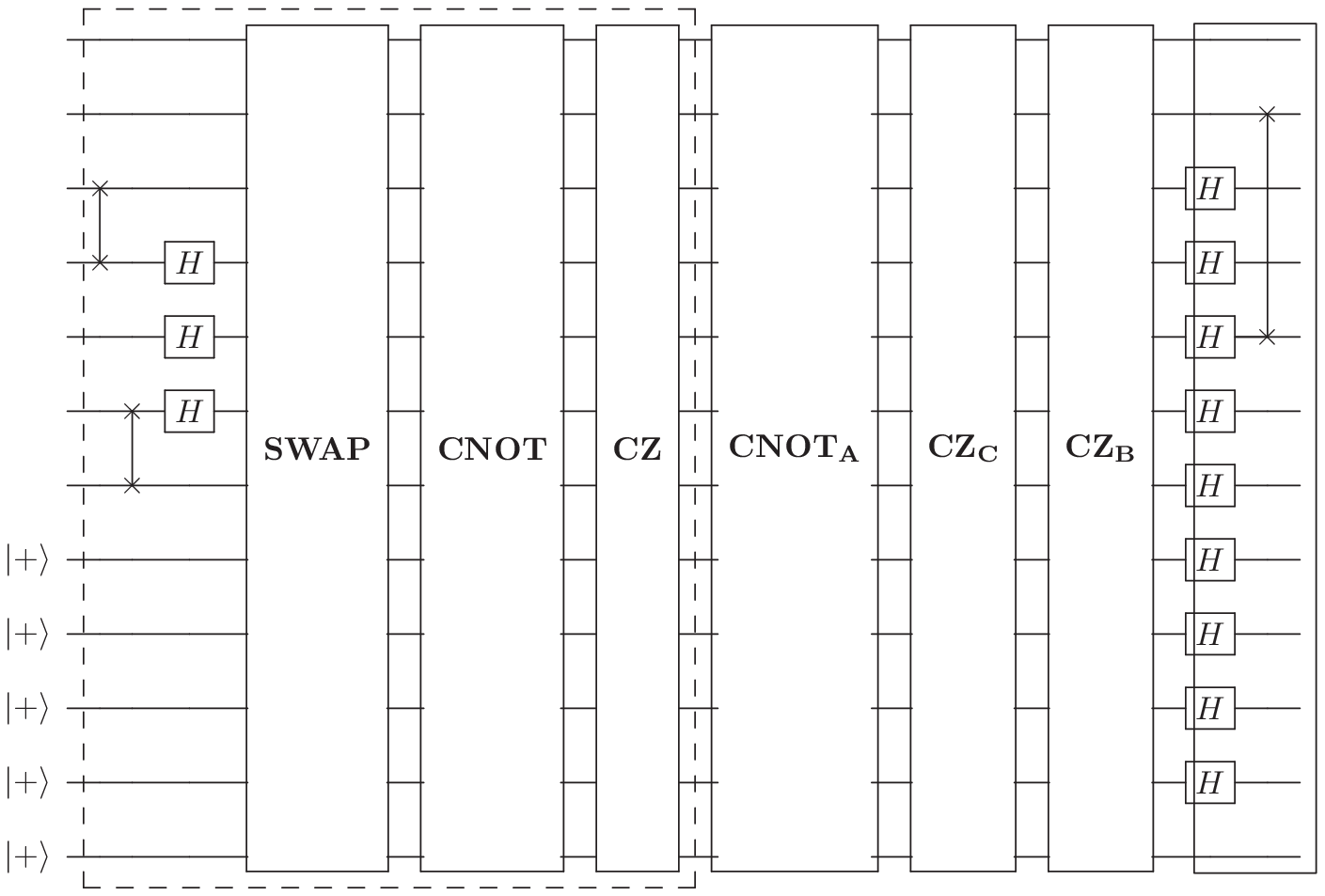,scale=0.8}
\fcaption{A fault tolerant quantum circuit to transform Steane code to $(3,4)$-QPC code. 
A quantum state at the most left side is a tensor product of a logical qubit of Steane code and ancilla qubits $|+\rangle^{\otimes 5}$, and a state at the other side is a logical qubit of $(3,4)$-QPC code. The box of the dotted line represents the Clifford operations to find the IABC form of the augmented Steane code, and that of the solid line indicates the inverse operation of the Clifford operations to find the IABC form of $(3,4)$-QPC code.}
\label{fig:steane_to_qpc}
\end{figure}

In this section, we show a conversion between $[[7,1,3]]$ Steane code and $(3,4)$-QPC code.
$(3,4)$-QPC code corresponds to $[[12,1,3]]$ stabilizer code.
Eleven stabilizer generators of the $(3,4)$-QPC code are defined as follows~\cite{Muralidharan:2014gk},
\begin{equation}
\mathcal{G}^{(3,4)-QPC} = \left(
\begin{array}{c}
Z_1Z_2,~Z_2Z_3,~Z_3Z_4, \\
Z_5Z_6,~Z_6Z_7,~Z_7Z_8, \\
Z_9Z_{10},~Z_{10}Z_{11},~Z_{11}Z_{12}, \\
X_1X_5X_2X_6X_3X_7X_4X_8, \\
X_5X_9X_6X_{10}X_7X_{11}X_8X_{12}
\end{array}
\right).
\end{equation}
For the conversion from Steane code to $(3,4)$-QPC code five ancilla qubits are added to Steane code, and no ancilla qubit is necessary for the QPC code for the opposite direction,
\begin{equation}
\mathcal{C}^{Steane}\otimes |+\rangle^{\otimes 5}\Leftrightarrow \mathcal{C}^{(3,4)-QPC}.
\end{equation}
All the Clifford gates are listed in Tab.~\ref{tab:clifford_list_steane_to_qpc}, and the corresponding circuit is shown in Fig.~\ref{fig:steane_to_qpc}.

\begin{sidewaystable}[t]
\centering
\tcaption{The list of the Clifford gates for the conversion from Steane code to $(3,4)$-QPC code.
All the gates have to be applied in the named order.}
\begin{tabular}{|c||c|} \hline 
Purpose & Gates \\ \hline 
\multirow{2}{*}{$\mathcal{G}^{Steane'}_{IABC}$} & SWAP(3,4), SWAP(6,7), H(4), H(5), H(6), $\mathbf{SWAP}$ = \{SWAP($i,i+1$) for $i=7\cdot 11$\}\\
& $\mathbf{CNOT}$=\{CNOT($i,12$) for $i=7\cdots 11$\}, $\mathbf{CZ}$=\{CZ($i, 12$) for $i=7\cdots 11$\}\\ \hline

$\mathcal{G}^{QPC}_{IABC}$ &  SWAP(2,5), \{H($i$) for $i=3\cdots 11$\} \\  \hline

\multirow{2}{*}{$A'_{St}\rightarrow A'_{QPC}$}  	& $\mathbf{CNOT_A}$ = \{CNOT(1,12), CNOT(3,12), CNOT(7,12), CNOT(5,12), CNOT(8,12), \\ 
									& CNOT(9,12), CNOT(10,12), CNOT(11,12)\},\\ \hline

$C'_{St}\rightarrow C'_{QPC}$ & 			$\mathbf{CZ_C}$ = \{CZ(4,12), CZ(7,12), CZ(8,12)\} \\ \hline

\multirow{3}{*}{$B''_{St} \rightarrow B'_{QPC}$} & $\mathbf{CZ_B}$=\{CZ(1,3), CZ(1,4), CZ(1,5), CZ(1,6), CZ(1,7), CZ(1, 8), CZ(1,9),  \\ 
									& CZ(1,11), CZ(2,7), CZ(2,5), CZ(2,8), CZ(3,5), CZ(3,6), CZ(3,7),   \\ 
									& CZ(3,8), CZ(3,9), CZ(3,10), CZ(3,11), CZ(5,8), CZ(5,9), CZ(5,10), CZ(5,11)\}\\ \hline
\end{tabular}
\label{tab:clifford_list_steane_to_qpc}
\end{sidewaystable}

%
%

\section{Discussion}\label{sec:discussion}

We have described a fault-tolerant code conversion scheme based on a binary picture of stabilizer code and Clifford gate.
The scheme first puts two stabilizer codes into specific forms we called a IABC form, and proceeds the conversion by applying a series of two-qubit Clifford gates.
Each gate is chosen from comparisons between the IABC forms of both codes.
The fault tolerance of the scheme can be kept by succeeding in quantum error correction during the entire code conversion process.
As case studies, we presented three conversions: the $[[5,1,3]]$ stabilizer code and Steane code, Steane code and $[[15,1,3]]$ Reed-Muller code and Steane code and $(3,4)$-QPC code.
Each code has its own advantage.
All the middle codes during the conversions described in the present work have minimum distance 3, and all the possible two-qubit errors have unique syndrome or are degenerate with an one-qubit error.

We believe that a code conversion has various applications in a quantum computing and a quantum communication.
The related previous works~\cite{Hill:2013uj,Anderson:2014ja} were motivated to increase the efficiency to store quantum information in a memory of finite capacity and to implement a universal fault-tolerant quantum computing.
In addition, a code conversion can be applicable to an interface between a computing component and a communication channel.
For example, a photon is a significant quantum system for a quantum communication.
As is well known, a photon loss is a dominating quantum error in a quantum communication.
In this regard, to date several quantum error correcting codes specialized for a photon loss have been developed~\cite{Ralph:2005ie,Grassl:1997ja}.

On the other hand, inside a quantum computer, most quantum errors are ones caused by operations of quantum gates, and reduced to a bit flip and a phase flip.
Besides, various quantum mechanical systems such as trapped ion, superconductor, semiconductor and so on are being considered to build a practical quantum computer.
It is also believed that several kinds of quantum systems have to be combined for a quantum computer.
Therefore, it is reasonable to implement an interface that fault tolerantly connects different quantum information processing components where one quantum error correcting code is more useful than the others.
We hope that the finding of the present work contributes to the research and development on such quantum interface systems.

\nonumsection{Acknowledgements}
\noindent
We thank David Poulin for useful comments and discussions.
This research was supported by the MSIP (Ministry of Science, ICT and Future Planning), Korea, under the ITRC (Information Technology Research Center) support program (IITP-2015-R0992-15-1017) supervised by the IITP (Institute for Information \& communication Technology Promotion) and the ICT R\&D program of MSIP/IITP (R0190-15-2030).

\nonumsection{Reference}

\bibliography{reference}
\bibliographystyle{unsrt}

\end{document}